\documentclass[12pt]{article}
\usepackage{hyperref}
\usepackage{amsmath}
\usepackage{graphicx}
\usepackage{amssymb} 
\usepackage{color}
\definecolor{mypurple}{rgb}{0.71,0.02,1}
\definecolor{myblue}{rgb}{0.14,0.11,0.49}
\newcommand{\Couleur}[1]{\textcolor{myblue}{#1}}
\newcommand{\Mat}[1]{{{\boldsymbol{#1}}}}

\def\be{\begin{equation}}
\def\ee{\end{equation}}
\def\bea{\begin{eqnarray}}
\def\eea{\end{eqnarray}}
\def\bi{\begin{itemize}}
\def\ei{\end{itemize}}
\def\dd{\mathrm{d}}

\date{}

\title{Classical-quantum correspondence and\\ wave packet solutions of the Dirac equation\\ in a curved spacetime}

\author{
Mayeul Arminjon\,$^{1}$ and Frank Reifler\,$^2$\\
$^1$ \small\it Laboratory ``Soils, Solids, Structures, Risks'', 3SR \\
\small\it (CNRS and Universit\'es de Grenoble: UJF, G-INP),\\
\small\it BP 53, F-38041 Grenoble cedex 9, France.\\
\small\it $^2$ Lockheed Martin Corporation, MS2 137-205,\\ 
\small\it 199 Borton Landing Road, Moorestown, New Jersey 08057, USA.
} 

\date{}

\begin{document}
\maketitle

\begin{abstract}
\noindent The idea of wave mechanics leads naturally to assume the well-known relation $E=\hbar \omega $ in the specific form $H=\hbar W$, where $H$ is the classical Hamiltonian of a particle and $W$ is the dispersion relation of the sought-for wave equation. We derive the expression of $H$ in a curved spacetime with an electromagnetic field. Then we derive the Dirac equation from factorizing the polynomial dispersion equation corresponding with $H$. Conversely, summarizing a recent work, we implement the geometrical optics approximation into a canonical form of the Dirac Lagrangian. Euler-Lagrange equations are thus obtained for the amplitude and phase of the wave function. From them, one is led to define a 4-velocity field which obeys exactly the classical equation of motion. The complete de Broglie relations are then derived exact equations.\\

\end{abstract}

  \section{Introduction}
\subsection{Context of this work}

The long-standing problem of quantum gravity may mean, of course, that we should try to better understand gravity and the quantum. More specifically, it may mean that we should try to better understand the transition between the classical and the quantum, especially in a curved spacetime. Quantum effects in the classical gravitational field are indeed being observed on neutral particles such as neutrons \cite{COW1975,WernerStaudenmannColella1979,Nesvizhevsky2002} or atoms \cite{RiehleBorde1991,KasevichChu1991}, with the neutrons being spin $\frac{1}{2}$ particles. This together motivates investigating the ``classical-quantum correspondence"---the correspondence between a classical Hamiltonian and a quantum wave equation---for the Dirac equation in a curved spacetime. 

\subsection{Foregoing results}

In a past work \cite{A22}, the classical-quantum correspondence was analyzed generally from an exact mathematical correspondence, observed by Whitham \cite{Whitham}, between a linear wave operator and its dispersion polynomial, and from the de Broglie-Schr\"odinger idea according to which a classical Hamiltonian describes the skeleton of a wave pattern. This analysis led later to deriving the Dirac equation from the classical Hamiltonian of a relativistic test particle in an electromagnetic field {\it or} in a curved spacetime \cite{A37,A39}. In the latter case, this derivation led to two {\it alternative} Dirac equations, in which the Dirac wave function is a complex {\it four-vector}, with the set of the components of the Dirac matrices building a $(2 \ 1)$ tensor \cite{A37,A39}. (This transformation behaviour may be designated by the acronym ``TRD": tensor representation of the Dirac fields.) In order to see if that makes sense physically, the quantum mechanics associated with the Dirac equation was then investigated in detail.\\

 First, it was found \cite{A40} that in a {\it Minkowski spacetime} in {\it Cartesian}  coordinates, the quantum mechanics of the original Dirac equation is {\it the same} whether, on a coordinate change, the wave function is transformed as a spinor and the Dirac matrices are left invariant (which, as is well known, is the standard transformation for this case), or if alternatively the TRD transformation mode is used. Moreover, the way in which this equivalence was obtained \cite{A40} makes it obvious that this equivalence holds also with the third transformation mode which can be considered for the Dirac equation, for which the wave function is left invariant and the set of the four Dirac matrices is transformed as a four-vector. (This is the transformation mode for the standard Dirac equation in  a gravitational field \cite{BrillWheeler1957+Corr, deOliveiraTiomno1962, ChapmanLeiter1976}.) Then it was found that also in a {\it general spacetime}, the standard equation and the two alternative equations based on TRD \cite{A39} behave similarly: e.g. the same hermiticity condition for the Hamiltonian holds for these equations \cite{A42}, and similar {\it non-uniqueness problems} of the Hamiltonian theory occur \cite{A42,A43}.

\subsection{Outline of this work}

In this conference paper, we intend to summarize the main part of the recent work \cite{A46}, and to present a few additional results. Those belong to Section \ref{classical-to-quantum}, which extends the former derivation of the Dirac equation from the classical Hamiltonian of a relativistic test particle \cite{A37,A39} to the situation with an electromagnetic field {\it and} in a curved spacetime. Then, summarizing the main results of Ref. \cite{A46}, Section \ref{quantum-to-classical} will go conversely from the Dirac equation to the classical motion through the geometrical optics approximation.
 
\section{From classical motion to Dirac equation}\label{classical-to-quantum}

  \subsection{Dispersion equation of a wave equation}
	
Consider a wave equation which is a linear PDE of the second order:
\be\label{linear 2nd order PDE}
\Couleur{\mathrm{P}\psi \equiv \ a_0(X)\Psi +a_1^\mu (X)\partial _\mu \Psi +a_2^{\mu \nu } (X)\partial _\mu \partial _\nu\Psi=0,}
\ee 
where \Couleur{$X $} is the position in the space-time, or more generally in the extended configuration manifold \Couleur{V} of a system of particles. [\Couleur{V} has dimension \Couleur{$N+1$}, where \Couleur{$N$} is the dimension of the configuration manifold.] To be more precise, Eq. (\ref{linear 2nd order PDE}) is the local expression of the intrinsic differential operator \Couleur{$\mathrm{P}$} (which acts on smooth sections \Couleur{$\psi$} of some vector bundle \Couleur{E} with base  \Couleur{V}) in a local chart \Couleur{$\chi : X\mapsto (x^\mu )$} and in a local frame field \Couleur{$(e_a)$} on \Couleur{E}, with \Couleur{$\Psi =(\Psi ^a)$} the column matrix made with the components of \Couleur{$\psi$} in the frame field \Couleur{$(e_a)$}, such that \Couleur{$\psi  =\Psi^a e_a$} in the domain of \Couleur{$(e_a)$}. The time coordinate is \Couleur{$t\equiv x^0/c$}.\\

\vspace{1mm}		
Let us look for ``locally plane-wave" solutions: \Couleur{$\ \Psi (X) = A\,\exp[ i\theta (X)]$}, with, at the point \Couleur{$X_0 \in \mathrm{V}$} considered,\Couleur{$\quad \partial_\nu K_\mu (X_0)=0$}, where \Couleur{$ K_\mu \equiv \partial _\mu \theta$} are the components of the wave covector. Note that
\Couleur{$K_0=-\omega/c$}, where \Couleur{$\omega  \equiv -\partial _t\theta$} is the angular frequency, and that \Couleur{${\bf k}\cong (K_j)$} is the spatial wave covector. (A Latin index \Couleur{$j=1,...,N$}, while a Greek one \Couleur{$\mu =0,...,N$}.)\\

\vspace{1mm}	
This leads \cite{A22,A37} to the {\it dispersion equation:}  
\be\label{Pi_X}
\Couleur{\Pi_X({\bf K})\equiv a_0(X) +i\,a_1^\mu (X)K_\mu +i^2a_2^{\mu \nu } (X)K_\mu K_\nu  =0.} 
\ee
Substituting $\Couleur{K_\mu  \hookrightarrow  \partial _\mu/i}\ $ determines the linear operator $\Couleur{\mathrm{P}}$ {\it uniquely} from the function $\Couleur{(X,{\bf K}) \mapsto \Pi_X({\bf K})}$, which is polynomial in $\Couleur{{\bf K}}$ at fixed $\Couleur{X}$ \cite{Whitham,A22}. This is also true \cite{A37,A39} in the case that \Couleur{$\Psi(X) \in {\sf C}^m$} with \Couleur{$m > 1$} in Eq. (\ref{linear 2nd order PDE}), in which case the coefficients common to \Couleur{$\mathrm{P}$} and \Couleur{$\Pi_X$} are \Couleur{$m \times m$} matrices \cite{A37}, so that \Couleur{$\Pi_X({\bf K})$} is then an \Couleur{$m \times m$} matrix, too.

\subsection{The classical-quantum correspondence} 

For any fixed \Couleur{$X \in \mathrm{V}$}, consider the dispersion equation (\ref{Pi_X}): \Couleur{$\ \Pi_X({\bf K}) =0$}, here assumed scalar (\Couleur{$m=1$}). This is a polynomial equation for \Couleur{$\quad \omega\equiv -cK_0$}. By following smoothly as a function of \Couleur{$X \in \mathrm{V}$} a particular root, assumed real, of this equation, we get a {\it dispersion relation(s)}: \Couleur{$\ \omega =W({\bf k};X)$}. (The existence of a such real root depending smoothly on \Couleur{$X \in \mathrm{V}$} is equivalent to the existence of a definite wave mode for the PDE (\ref{linear 2nd order PDE}) \cite{Whitham}.) As shown by Whitham \cite{Whitham} (see also Ref. \cite{A22}), the propagation of \Couleur{$\ {\bf k}$} obeys a {\it Hamiltonian system:}

\be \label{Hamilton-W-k} 
\Couleur{\frac{\dd K_j}{\dd t}= -\frac{\partial  W}{\partial x^j}},\qquad \Couleur{\frac{\dd x^j}{\dd t}= \frac{\partial  W}{\partial K_j}}\qquad (\Couleur{j=1,...,N}).
\ee

\vspace{3mm}
On the other hand, according to the {\it wave mechanics} inaugurated by de Broglie and Schr\"odinger, a classical Hamiltonian \Couleur{$H({\bf p},{\bf x},t)=H({\bf p};X)$} describes the skeleton of a wave pattern. Then, the wave equation should give a dispersion \Couleur{$W$} with the same Hamiltonian trajectories as \Couleur{$H$}. The simplest way to get that is to {\it assume} that \Couleur{$H$} and \Couleur{$W$} are proportional, \Couleur{$H=\hbar W$}... This leads first to the de Broglie relations in traditional form: \Couleur{$ E=\hbar \omega$} and \Couleur{$ \quad {\bf p}=\hbar {\bf k}$}. Then, substituting \Couleur{$K_\mu  \hookrightarrow  \partial _\mu/i$}, it leads to the correspondence between a classical Hamiltonian and a wave operator. See Refs. \cite{A22,A37} for details. Thus, setting 
\be\label{P_mu from p_j and H}
\Couleur{P_j\equiv p_j \quad (j=1,...,N)}\quad \mathrm{and}\ \ \Couleur{P_0\equiv -\frac{H}{c}},
\ee
we get the de Broglie relations in a condensed form:
\be\label{de Broglie}
\Couleur{P_\mu =\hbar K_\mu} \qquad(\Couleur{\mu =0,...,N}).
\ee


\subsection{The classical-quantum correspondence needs using a connection}

The dispersion polynomial $\Couleur{\Pi _X({\bf K})}$ and the condition $\Couleur{\partial_\nu K_\mu (X)=0}$ stay invariant only inside any class of coordinate systems connected by ``infinitesimally-linear" changes \cite{A22}, i.e., ones such at the point $\Couleur{X((x_0^\mu))=X((x_0'^\rho))}$ considered: 
\be \label{infinit-linear}
\Couleur{\frac{\partial ^2x'^\rho}{\partial x^\mu\partial x^\nu}=0, \qquad \mu ,\nu ,\rho  \in \{0,...,N\}}.
\ee

\vspace{5mm}
\noindent One example \cite{A22,A37} of a such class is constituted by the locally-geodesic coordinate systems at $\Couleur{X}$ for a pseudo-Riemannian metric $\Couleur{\Mat{g}}$ on  $\Couleur{\mathrm{V}}$, {\it i.e.}, 
\be\label{LGCS}
\Couleur{g_{\mu \nu,\rho}(X)=0, \qquad \mu,\nu, \rho \in \{0,...,N\}}. 
\ee
\vspace{1mm}
Specifying, at each point \Couleur{$X \in \mathrm{V}$}, a class \Couleur{$\mathcal{C}_X$} of coordinate systems valid in a neighborhood of \Couleur{$X$}, any two of which exchange by a transition map satisfying (\ref{infinit-linear}), is exactly equivalent to choosing a {\it torsionless connection} \Couleur{$D$} on the tangent bundle \Couleur{$\mathrm{TV}$} \cite{A39}. Given any such connection, one substitutes \Couleur{$\partial _\mu \hookrightarrow D_\mu $} into the wave equation (\ref{linear 2nd order PDE}), into the local plane-wave condition which rewrites thus \Couleur{$\quad D_\nu K_\mu (X_0)=0$}, and into the correspondence from the dispersion equation (\ref{Pi_X}) to the wave equation. That correspondence writes thus \Couleur{$K_\mu  \hookrightarrow  D _\mu/i$}. It applies also to the case where \Couleur{$\Psi(X) \in {\sf C}^m$} with \Couleur{$m > 1$}, provided that the dispersion equation (\ref{Pi_X}) and the wave equation are {\it first-order} in fact, i.e. \Couleur{$a_2^{\mu \nu }=0$} \cite{A39}.

 \subsection{Hamiltonian of a particle in a curved spacetime}
In a curved spacetime $\Couleur{(\mathrm{V},\Mat{g})}$ with an electromagnetic field of 4-potential $\Couleur{V_\mu }$, the world line of a test particle corresponds with an extremum of the generally-covariant action integral
\be
\Couleur{S=\int -mc \,ds-\frac{e}{c}V_\mu dx^\mu},
\ee
where 
\be\label{ds^2}
\Couleur{ds^2=g_{\lambda\rho}dx^\lambda dx^\rho}.
\ee 
It follows that the motion derives from the following Lagrangian:
\be\label{Extended L}
\Couleur{\mathcal{L}\left(x^\mu, u'^\nu \right) = -mc \sqrt{g_{\mu \nu }u'^\mu u'^\nu }-(e/c)V_\mu u'^\mu , \quad u'^\nu \equiv dx^\nu/d\xi }, 
\ee
with \Couleur{$\xi$} an arbitrary parameter along the world line of the particle. The canonical momenta derived from this Lagrangian are
\be\label{P_mu from extended L}
\Couleur{P_\mu \equiv \partial\mathcal{L}/\partial u'^\mu =-mc \frac{u'_\mu}{\sqrt{g_{\lambda\rho}u'^\lambda u'^\rho}}-(e/c)V_\mu} .
\ee
The Lagrangian (\ref{Extended L}) is an ``extended Lagrangian" in the sense of Johns \cite{Johns2005}. As shown by Johns \cite{Johns2005} (Sect. 11.9), we may associate with an extended Lagrangian like $\Couleur{\mathcal{L}}$ a ``traditional Lagrangian" $\Couleur{L}$, by setting
\be
\Couleur{L(x^j, \frac{dx^k}{dx^0}, x^0)\equiv \frac{d\xi}{dx^0} \mathcal{L}(x^\mu,\frac{dx^\nu}{d\xi})}.
\ee
From the latter, we deduce by Legendre transform a ``traditional Hamiltonian" \Couleur{$H'(p_j, x^k, x^0)$}. The  ``traditional momenta" \Couleur{$p_j$} coincide with the corresponding ``extended momenta" \Couleur{$P_\mu$} (for \Couleur{$\mu =1,...,N$}), the latter ones being canonically derived from the extended Lagrangian $\Couleur{\mathcal{L}}$ by Eq. (\ref{P_mu from extended L})$_1$ (Ref. \cite{Johns2005}, Eq. (11.12)). That is, we have 
\be\label{p_j=P_j}
\Couleur{p_j=P_j \qquad (j=1,...,N)}. 
\ee
The traditional Hamiltonian is simply (Ref. \cite{Johns2005}, Eq. (11.14))
\be\label{Traditional H}
\Couleur{H'(p_j, x^k, x^0)=-P_0(x^\mu,u'^\nu )} .
\ee

\vspace{3mm}
At this stage, we can specialize the parameter \Couleur{$\xi $} to be the four-length \Couleur{$s$}, Eq. (\ref{ds^2}). In that case, the vector with components \Couleur{$u'^\mu $} is the four-velocity, \Couleur{$u'^\mu =u^\mu \equiv dx^\mu/ds$}. From (\ref{ds^2}), it verifies \Couleur{$g^{\mu \nu} u_\mu u_\nu =1$}, as is well known. Hence, with \Couleur{$\xi =s$} the canonical momenta (\ref{P_mu from extended L}) become:
\be\label{Canonical momenta}
\Couleur{P_\mu  =-mc u_\mu - (e/c)V_\mu}.
\ee
Again because \Couleur{$g^{\mu \nu} u_\mu u_\nu =1$}, they verify the following energy equation:
\be\label{Energy equation}
\Couleur{g^{\mu \nu} \left(P_\mu +\frac{e}{c}V_\mu \right)\left(P_\nu+\frac{e}{c}V_\nu \right) -m^2 c^2=0}.
\ee
In the expression (\ref{Traditional H}) of the traditional Hamiltonian $\Couleur{H'}$, the time coordinate \Couleur{$x^0$} is arbitrary. Let us get $\Couleur{H'}$ as function of the same momenta \Couleur{$p_j$} and the same space coordinates \Couleur{$x^k$}, but with the time coordinate \Couleur{$t\equiv x^0/c$}. We do that directly in Hamilton's equations for $\Couleur{H'}$. We must set:
\be\label{H=-cP_0}
\Couleur{H(p_j, x^k, t) =cH'(p_j, x^k, x^0)=-cP_0\left(x^\mu,u^\nu \right)}.
\ee
Note that Eqs. (\ref{p_j=P_j}) and (\ref{H=-cP_0}) are consistent with the definition (\ref{P_mu from p_j and H}).

 \subsection{A variant derivation of the Dirac equation}
\noindent The dispersion equation associated with the energy equation (\ref{Energy equation}) by wave mechanics, i.e., by the de Broglie relations (\ref{de Broglie}), is:
\be\label{Dispersion}
\Couleur{g^{\mu \nu} \left(\hbar K_\mu+\frac{e}{c}V_\mu \right) \left(\hbar K_\nu +\frac{e}{c}V_\nu \right)-m^2 c^2= 0} .
\ee
Applying directly the correspondence \Couleur{$\ K_\mu  \hookrightarrow  D_\mu/i\ $} to the dispersion equation (\ref{Dispersion}), leads to a specific form of the curved-spacetime Klein-Gordon equation \cite{A39}. Instead, one may try a {\it factorization} with matrix coefficient \Couleur{$\alpha,\gamma ^\mu$}, etc.:
\bea\nonumber
\Couleur{\Pi _X({\bf K})} & \equiv & \Couleur{\left[g^{\mu \nu} \left(K_\mu+eV_\mu \right) \left( K_\nu +eV_\nu \right)-m^2 \right ]{\bf 1}}\\ \label{Factorize Pi}
& =? & \Couleur{(\alpha +i \gamma ^\mu K_\mu)(\beta +i \zeta ^\nu K_\nu)}.\qquad (\Couleur{\hbar =1=c})
\eea

\vspace{2mm}
Identifying the coefficients (with noncommutative algebra), and substituting \Couleur{$K_\mu  \hookrightarrow  D _\mu/i$}, leads to the Dirac equation \cite{A37,A39}:
\be \label{Dirac-brut}
\Couleur{(i\gamma^\mu \, (D _\mu+ieV_\mu ) -m) \psi=0},
\ee
with the anticommutation relation, resulting from the identification in Eq. (\ref{Factorize Pi}):
\be\label{Clifford}
\Couleur{\gamma ^\mu \gamma ^\nu + \gamma ^\nu \gamma ^\mu = 2g^{\mu \nu}\,{\bf 1}}.
\ee

\vspace{4mm}

\section{From Dirac equation to classical motion}\label{quantum-to-classical}

\subsection{General Dirac Lagrangian in a curved spacetime}

The following Lagrangian (density) \cite{A43} generalizes the ``Dirac Lagrangian" (e.g. \cite{BrillWheeler1957+Corr,Leclerc2006}) valid for the standard Dirac equation in a curved spacetime:
\be\label{Lagrangian-density}
\Couleur{l=\sqrt{-g}\ \frac{i}{2}\left[\,\overline{\Psi}\gamma^{\mu}(D_{\mu}\Psi)-
\left(\overline{D_{\mu}\Psi} \right)\gamma^{\mu}\Psi+2im\overline{\Psi}\Psi\right]},
\ee
where \Couleur{$\overline{\Psi}\equiv \Psi ^\dagger A$}  is the generalized Dirac adjoint of \Couleur{$\Psi \equiv (\Psi^a)$}. The field \Couleur{$X\mapsto A(X)$} designates the {\it hermitizing matrix}, such that \Couleur{$A^\dagger =A,\ (A\gamma ^\mu )^\dagger=A\gamma ^\mu$} \cite{Pauli1936,A40}. (Here \Couleur{$M^\dagger\equiv (M^*)^T$} is the hermitian conjugate of a matrix \Couleur{$M$}.) In a curved spacetime, that matrix becomes indeed generally a field \cite{A42}. In the usual Dirac Lagrangian, the field \Couleur{$A$} is the constant matrix \Couleur{$\gamma ^{\natural 0}$}, where \Couleur{$(\gamma ^{\natural \alpha })$} is a set of constant ``flat" Dirac matrices, i.e., ones obeying the anticommutation relation (\ref{Clifford}) with the Minkowski metric \Couleur{$\eta ^{\alpha \beta }$}---provided the set \Couleur{$(\gamma ^{\natural \alpha })$} is chosen such that \Couleur{$\gamma ^{\natural 0}$} be a hermitizing matrix for that set.\\

\vspace{1mm}
The Euler-Lagrange equation for the Lagrangian (\ref{Lagrangian-density})  is the generalized Dirac equation \cite{A42,A43}: 
\be\label{Dirac-general}
\Couleur{\gamma ^\mu D_\mu\Psi=-im\Psi-\frac{1}{2}A^{-1}(D_\mu (A\gamma ^\mu) )\Psi}.
\ee
This coincides with the usual form iff \Couleur{$D_\mu (A\gamma ^\mu)=0$}. That is always the case  \cite{A42} for the standard Dirac equation in a curved spacetime (the ``Dirac-Fock-Weyl" or DFW equation). In Eqs. (\ref{Lagrangian-density}) and (\ref{Dirac-general}), the covariant derivatives \Couleur{$D_\mu $} correspond to an arbitrary connection \Couleur{$D $} defined on the complex vector bundle \Couleur{$\mathrm{E}$}, in which the Dirac wave function \Couleur{$\psi $} is living. 
\footnote{\
The connection \Couleur{$D $} and the covariant derivatives \Couleur{$D_\mu $}  extend as usual to the dual bundle \Couleur{$\mathrm{E}^\circ$} of \Couleur{$\mathrm{E}$}, and to tensor products such as \Couleur{$\mathrm{E} \otimes \mathrm{E}^\circ$}. Moreover, on a tensor product such as \Couleur{$\mathrm{TV }\otimes \mathrm{E} \otimes \mathrm{E}^\circ$} (which is the relevant bundle for the Dirac matrices \cite{A45}), the relevant connection is got from considering the Levi-Civita connection on the component bundle \Couleur{$\mathrm{TV }$} \cite{A42,A45}.
}
That vector bundle is assigned to be a ``spinor bundle", i.e. essentially, one for which it exists a global field \Couleur{$\gamma $} of Dirac matrices, consistent with the anticommutation relation (\ref{Clifford}). See Ref. \cite{A45} for details.


\subsection{Local similarity (or gauge) transformations}

Admissible coefficient fields \Couleur{$(\gamma^\mu ,A)$} for the general Dirac equation (\ref{Dirac-general}) are ones such that the anticommutation relation (\ref{Clifford}) is satisfied and that the field \Couleur{$A$} is hermitizing. Given any {\it local similarity transformation} \Couleur{$S:\ X \mapsto S(X) \in {\sf GL(4,C)}$}, depending smoothly on \Couleur{$X \in \mathrm{V}$}, other admissible coefficient fields are 
\be\label{similarity}
\Couleur{\widetilde{\gamma} ^\mu =  S^{-1}\gamma ^\mu S \quad (\mu =0,...,3), \qquad \widetilde {A} \equiv S^\dagger A S},
\ee
in the sense that the anticommutation relation (\ref{Clifford}) remains satisfied [in the same spacetime \Couleur{$(\mathrm{V},\Mat{g})$} !] with the new field of Dirac matrices \Couleur{$\widetilde{\gamma} ^\mu$}, and moreover the matrix field \Couleur{$\widetilde {A}$} is a hermitizing matrix field for \Couleur{$\widetilde{\gamma} ^\mu$} \cite{A42}.\\

\vspace{1mm}
The relevant Hilbert space scalar product is \cite{A42}
\be\label{scalar product}
\Couleur{(\Psi \mid \Phi ) \equiv \int \Psi ^\dagger A \gamma ^0 \Phi \sqrt{-g}\ d^3{\bf x}}. 
\ee
It transforms isometrically under the gauge transformation (\ref{similarity}), if one transforms the wave function according to \Couleur{$\widetilde{\Psi} \equiv S^{-1}\Psi$} \cite{A43}.\\

The Dirac equation (\ref{Dirac-general}) is covariant under the similarity (\ref{similarity}), if the connection matrices \Couleur{$\Gamma _\mu$}, such that \Couleur{$D_\mu =\partial_\mu +\Gamma _\mu$}, change thus \cite{ChapmanLeiter1976,A43}:
\be\label{Gamma-tilde-psitilde=S^-1 psi}
\Couleur{\widetilde{\Gamma }_\mu = S^{-1}\Gamma _\mu S+ S^{-1}(\partial _\mu S)}.
\ee
For the DFW equation, the gauge transformation (\ref{similarity}) is restricted to belong to the spin group: \Couleur{$\forall X \ S(X) \in {\sf Spin(1,3)}$}. Then, the relation (\ref{Gamma-tilde-psitilde=S^-1 psi}) is automatically satisfied \cite{A43}.


\subsection{Reduction of the Dirac equation to canonical form}

If \Couleur{$D_\mu (A\gamma ^\mu)=0$} {\it and} the \Couleur{$\Gamma _\mu $}'s are zero, the Dirac equation (\ref{Dirac-general}) writes
\be\label{Dirac-canonical}
\Couleur{\gamma ^\mu \partial _\mu\Psi=-im\Psi}.
\ee

\vspace{2mm}
\noindent {\bf Theorem 1 \cite{A46}.} {\it In the neighborhood of any event \Couleur{$X$}, the Dirac equation (\ref{Dirac-general}) can be put into the {\it canonical form} (\ref{Dirac-canonical}) by a local similarity transformation.}\\

\noindent Outline of the proof: i) By Theorem 3 of Ref. \cite{A42} [Sect. 3.4, Eq. (54)], a similarity \Couleur{$T$} defined by Eq. (\ref{similarity}) [with \Couleur{$T$} instead of \Couleur{$S$}] brings the general Dirac equation (\ref{Dirac-general}) to the ``normal" form (\Couleur{$D_\mu (A\gamma ^\mu)=0$}), iff
\be\label{general-to-normal}
\Couleur{A\gamma ^\mu D_\mu T=-(1/2) [D_\mu (A\gamma ^\mu )]T}.
\ee
ii) By Theorem 2 of Ref. \cite{A45} [Sect. 6.2, Eq. (65)], a similarity \Couleur{$S$} defined by Eq. (\ref{similarity}) and such that Eq. (\ref{Gamma-tilde-psitilde=S^-1 psi}) is satisfied, brings a normal Dirac equation to the canonical form (\ref{Dirac-canonical}), iff
\be\label{normal-to-canonical}
\Couleur{A\gamma ^\mu \partial _\mu S=-A\gamma ^\mu \Gamma _\mu S}.
\ee
iii) Due to the hermitizing character of matrix \Couleur{$A$}, and due to the fact that, by construction, the Hermitian matrix \Couleur{$A\gamma ^0$} which defines the scalar product (\ref{scalar product}) is positive definite \cite{A42}, both (\ref{general-to-normal}) and (\ref{normal-to-canonical}) are symmetric hyperbolic systems.$\hspace{118mm} \square$


\subsection{Geometrical optics approximation of Dirac Lagrangian}


The Lagrangian for the Dirac equation in an electromagnetic (e.m.) field is got by substituting \Couleur{$D_\mu \hookrightarrow D_\mu +\frac{ie}{\hbar c}V_\mu $}, where \Couleur{$e$} is the particle's electric charge. It follows that, for the canonical Dirac equation (\ref{Dirac-canonical}), the Lagrangian in an e.m. field is
{\setlength\arraycolsep{2pt}
\bea\label{Lagrangian-canonical}
\Couleur{l} & = & \Couleur{\sqrt{-g}\ \frac{i\hbar c}{2}\left[\Psi^\dagger A\gamma^{\mu}(\partial _{\mu}\Psi)-\left(\partial _{\mu}\Psi \right)^\dagger A\gamma^{\mu}\Psi+\frac{2imc}{\hbar }\Psi^\dagger A\Psi \right]-}{} 
\nonumber\\ 
  & & {}\Couleur{- \sqrt{-g}\,(e/c)J^\mu V_\mu },
\eea}
with \Couleur{$\nabla _\mu (A\gamma ^\mu )=0$}, where \Couleur{$\nabla _\mu$} is the covariant derivative obtained on the relevant tensor product bundle (see Footnote 1) if the connection matrices \Couleur{$\Gamma _\mu$} of the connection \Couleur{$D$}  are zero when \Couleur{$D$} is acting specifically on the bundle \Couleur{E} \cite{A46}.  To implement the geometrical optics approximation following Whitham \cite{Whitham}, we set first 
\be\label{Psi =chi e^{i theta }}
\Couleur{\Psi =\chi e^{i \theta }},
\ee
where \Couleur{$\chi  = \chi  (X )$} is also a complex wave function with four components, and \Couleur{$\theta  = \theta  (X )$} is a real phase. We assume then that \Couleur{$\chi $} is slowly changing compared to the rapidly changing phase \Couleur{$\theta $}. That is, the geometrical optics approximation consists in assuming that
\be
 \Couleur{\partial _\mu \chi \ll (\partial _\mu \theta )\chi }. 
\ee
Substituting (\ref{Psi =chi e^{i theta }}) into the Lagrangian (\ref{Lagrangian-canonical}) with this approximation, yields
\be\label{Lagrangian-Whitham}
\Couleur{l'=c\sqrt{-g}\left[\left (-\hbar \partial _\mu \theta -\frac{e}{c}V_\mu  \right )\chi ^\dagger  A\gamma^{\mu}\chi -mc\chi ^\dagger A\chi \right]}.
\ee

\vspace{1mm}
The Euler-Lagrange equations are then \cite{A46}:
\bea\label{Euler-Lagrange-Whitham}
\Couleur{\left (-\hbar \partial _\mu \theta -\frac{e}{c}V_\mu  \right ) A\gamma^{\mu}\chi = mcA\chi },\\
\Couleur{\partial _\mu \left (c\sqrt{-g}\ \chi ^\dagger A\gamma^{\mu}\chi \right )= 0}.
\eea


\subsection{Classical trajectories}


{\bf Theorem 2 \cite{A46}.} {\it From \Couleur{$\Psi =\chi e^{i\theta }$}, define a four-vector field \Couleur{$u^\mu $} and a scalar field \Couleur{$J$} thus:}
\bea\label{Def u_mu}
\Couleur{u_\mu \equiv -\frac{\hbar }{mc}\partial _\mu \theta -\frac{e }{mc^2}V_\mu},\\
\Couleur{u^\mu \equiv g^{\mu \nu }\,u_\nu },\\
\Couleur{J \equiv  c\ \chi ^\dagger A\chi}.
\eea
{\it Then the Euler-Lagrange equations (\ref{Euler-Lagrange-Whitham}) imply}
\bea
\Couleur{\nabla _\mu (J u^\mu )=0},\\
\Couleur{g^{\mu \nu }\,u_\mu u_\nu =1},\\
\Couleur{\nabla _\mu u_\nu -\nabla _\nu u_\mu =-(e/mc^2)\,F_{\mu \nu }}.
\eea
{\it The first equation is the conservation of the probability current. The two last equations imply the classical equation of motion for a test particle in an electromagnetic field in a curved spacetime.}

\subsection{De Broglie relations}


The canonical momenta of a {\it classical particle} are given by Eq. (\ref{Canonical momenta}):
\be\label{Canonical momenta-bis}
\Couleur{P_\mu =-mc u_\mu-(e/c)V_\mu} .
\ee

\vspace{1.5mm}
On the other hand, following a {\it Dirac quantum particle} in the geometrical approximation, we were led to define a 4-velocity field \Couleur{$u^\mu$} from the phase \Couleur{$\theta $} of the wave function, Eq. (\ref{Def u_mu}):
\be\label{Def u_mu-bis}
\Couleur{u_\mu \equiv -\frac{\hbar }{mc}\partial _\mu \theta -\frac{e }{mc^2}V_\mu}.
\ee
We saw that the field \Couleur{$u^\mu$} obeys exactly the classical equations of motion, which are Hamiltonian equations for which the canonical momenta are given by Eq. (\ref{Canonical momenta-bis}). But, remembering the definition \Couleur{$K_\mu \equiv \partial _\mu \theta$} of the wave covector from the phase \Couleur{$\theta $} of the wave function, we rewrite Eq. (\ref{Def u_mu-bis}) as
\be
\Couleur{-mc u_\mu-(e/c)V_\mu \equiv \hbar K _\mu}.
\ee
That is, we derive exactly the de Broglie relations (\ref{de Broglie}).


\section{Conclusion}


The Dirac equation {\it in a curved spacetime with electromagnetic field} may be ``derived'' from the classical Hamiltonian \Couleur{$H$} of a relativistic test particle. One has to {\it postulate} \Couleur{$H=\hbar W$} (i.e., \Couleur{$E=\hbar \omega $}), where \Couleur{$W$} is the dispersion relation of the sought-for wave equation. Then one factorizes the obtained dispersion polynomial.   \\

Conversely, to describe ``wave packet'' motion, we implemented the geometrical optics approximation into a canonical form of the Dirac Lagrangian. From the equations obtained thus for the amplitude and phase of the wave function, one is led to define a 4-velocity field. This obeys exactly the classical equation of motion.\\

The de Broglie relations $\Couleur{P_\mu =\hbar K_\mu}$ are then {\it derived} exact equations. 

\vspace{2mm}


\begin{thebibliography}{9}
\small

\bibitem{A22} 
Arminjon M., {\it On the relation Hamiltonian -- wave equation, and on non-spreading solutions of Schr\"odinger's equation}, Nuovo Cim. {\bf 114B} (1999) 71--86.

\bibitem{A37}
Arminjon M., {\it Dirac equation from the Hamiltonian and the case with a gravitational field}, Found. Phys. Lett. {\bf 19} (2006) 225--247. \href{http://arxiv.org/abs/gr-qc/0512046}{[arXiv:gr-qc/0512046]}

\bibitem{A39}
Arminjon M., {\it Dirac-type equations in a gravitational field, with vector wave function}, Found. Phys. {\bf 38} (2008) 1020--1045. \href{http://arxiv.org/abs/gr-qc/0702048}{[arXiv:gr-qc/0702048]}

\bibitem{A40}
Arminjon M. and Reifler F., {\it Dirac equation: Representation independence and tensor transformation}, Braz. J. Phys. {\bf 38} (2008) 248--258. \href{http://arxiv.org/abs/0707.1829}{[arXiv:0707.1829 (quant-ph)]}

\bibitem{A42}
Arminjon M. and Reifler F., {\it Basic quantum mechanics for three Dirac equations in a curved spacetime}, Braz. J. Phys. {\bf 40} (2010) 242--255.[\href{http://arxiv.org/abs/0807.0570}{arXiv:0807.0570 (gr-qc)}]

\bibitem{A43}
Arminjon M. and Reifler F., {\it A non-uniqueness problem of the Dirac theory in a curved spacetime}, Ann. Phys. (Berlin) {\bf 523} (2011) 531--551. \href{http://arxiv.org/abs/0905.3686}{[arXiv:0905.3686 (gr-qc)]}

\bibitem{A45}
Arminjon M. and Reifler F., {\it Four-vector vs. four-scalar representation of the Dirac wave function}, \href{http://arxiv.org/abs/1012.2327}{arXiv:1012.2327 (gr-qc)}.

\bibitem{A46}
Arminjon M. and Reifler F., {\it Generalized de Broglie relations for Dirac equations in curved spacetimes},  \href{http://arxiv.org/abs/arXiv:1103.3201}{arXiv:1103.3201 (gr-qc)}

\bibitem{BrillWheeler1957+Corr}
Brill D. R. and Wheeler J. A., {\it Interaction of neutrinos and gravitational fields}, Rev. Modern Phys. {\bf 29} (1957) 465--479. Erratum: Rev. Modern Phys. {\bf 33} (1961) 623--624.

\bibitem{ChapmanLeiter1976}
Chapman T. C. and Leiter D. J., {\it On the generally covariant Dirac equation}, Am. J. Phys. {\bf 44} (1976) 858--862.

\bibitem{COW1975}
Colella R., Overhauser A. W. and Werner S. A., {\it Observation of gravitationally induced quantum interference}, Phys. Rev. Lett. {\bf 34} (1975) 1472--1474.

\bibitem{Johns2005}
Johns O. D., {\it Analytic Mechanics for Relativity and Quantum Mechanics}, Oxford University Press 2005, pp. 267-271 and 563-565.

\bibitem{KasevichChu1991}
Kasevich M. and Chu S., {\it Atomic interferometry using stimulated Raman transitions}, Phys. Rev. Lett. {\bf 67} (1991) 181--184.

\bibitem{Leclerc2006}
Leclerc M., {\it Hermitian Dirac Hamiltonian in the time-dependent gravitational field}, {\it Class. Quant. Grav.} {\bf 23} (2006) 4013--4020. \href{http://arxiv.org/abs/gr-qc/0511060}{[arXiv:gr-qc/0511060]}

\bibitem{Nesvizhevsky2002} 
Nesvizhevsky V. V. {\it et al.}, {\it Quantum states of neutrons in the Earth's gravitational field}, Nature {\bf 415} (2002) 297--299.

\bibitem{deOliveiraTiomno1962}
de Oliveira C. G. and Tiomno J., {\it Representations of Dirac equation in general relativity}, Nuovo Cim. {\bf 24} (1962) 672--687.

\bibitem{Pauli1936}
Pauli W., {\it Contributions math\'ematiques \`a la th\'eorie des matrices de Dirac}, Ann. Inst. Henri Poincar\'e {\bf 6} (1936) 109--136).

\bibitem{RiehleBorde1991}
Riehle F., Kisters Th., Witte A., Helmcke J. and Bord\'e Ch. J., {\it Optical Ramsey spectroscopy in a rotating frame, Sagnac effect in a matter-wave interferometer}, Phys. Rev. Lett. {\bf 67} (1991) 177--180.

\bibitem{WernerStaudenmannColella1979}
Werner S. A., Staudenmann J. A. and Colella R., {\it Effect of Earth's rotation on the quantum mechanical phase of the neutron}, Phys. Rev. Lett. {\bf 42} (1979) 1103--1107.

\bibitem{Whitham} 
Whitham G. B., {\it Linear and Non-linear Waves}, J. Wiley \& Sons, New York 1974, Section 11: Linear dispersive waves.


\end{thebibliography}
\end{document}